# Drug Repurposing For SARS-COV-2 Using Molecular Docking

Imra Aqeel
Biomedical Informatics Research Lab, DCIS.
Pakistan Institute of Engineering and Applied Sciences, PIEAS
Islamabad, Pakistan
imraaqeel@pieas.edu.pk

Abdul Majid
Biomedical Informatics Research Lab, DCIS.
Pakistan Institute of Engineering and Applied Sciences, PIEAS
Islamabad, Pakistan
abdulmajid@pieas.edu.pk

Muhammad Ismail
Biomedical Informatics Research Lab, DCIS.
Pakistan Institute of Engineering and Applied Sciences, PIEAS
Islamabad, Pakistan
mismail_19@pieas.edu.pk

Hina Bashir
Department of Bio Sciences.
COMSATS Institute of Information Technology
Islamabad, Pakistan
fa18-bsi-016@isbstudent.comsats.edu.pk

*Abstract*— Drug repurposing is an unconventional approach that is used to investigate new therapeutic aids of existing and shelved drugs. Recent advancement in technologies and the availability of the data of genomics, proteomics, transcriptomics, etc., and with the accessibility of large and reliable database resources, there are abundantly of opportunities to discover drugs by drug repurposing in an efficient manner. The recent pandemic of SARS-COV-2, that caused the death of 6,245,750 human beings to date, has tremendously increase the exceptional usage of bioinformatics tools in interpreting the molecular characterizations of viral infections. In this paper, we have employed various bioinformatics tools such as AutoDock-Vina, PyMol, and Discovery Studio and found a leading drug candidate Cepharanthine (CEP) that has shown better results and effectiveness than recently used antiviral drug candidates such as Favipiravir, IDX184, Remdesivir, and Ribavirin. This paper has analyzed CEP's potential therapeutic importance as a drug of choice in managing COVID-19 cases. It is anticipated that proposed study would be beneficial for researchers and medical practitioners in handling SARS-CoV-2 and its variant related diseases.

## I. Introduction

Covids (CoVs) all around are the diversion for delicate to guaranteed respiratory plot infections. In the past various years, two essentially pathogenic CoVs, the very shocking respiratory issue Covid (SARS-CoV) and the Middle East respiratory condition Covid (MERS-CoV), both imparted from animals to people, set off by and large scourges, in 2003 and 2012, independently, with high passing rates [1], [2], [3], [4]. In December 2019, a Covid convincing sickness (named COVID-19) was seen in Wuhan, area of Hubei, China, accomplished by a new pathogenic CoV, named SARS-CoV-2. The pollution spread quickly from China to all nations, and on March eleventh, 2020, it was announced a pandemic by the World Health Organization (WHO) [1], [5], [6]. Notwithstanding mass immunization overall by crisis embraced antibodies like Pfizer-BioNTech, Janssen, and Moderna, COVID-19 tends to a danger to human thriving [7], [8], [9], [10].

At the hour of this composition the death pace of the SARS-CoV-2 is as of now assessed in the scope of 0.5-6 %; and keeping in mind that COVID-19 has all the earmarks of being less lethal than SARS (∼10 %) or MERS (∼40 %), it is by all accounts more infectious, with a regenerative number (Ro) in the reach 2.0-6.5, higher than SARS and MERS, which could clarify the speed of its spread [1].

Today, no therapeutics is accessible, and recurring flow sickness the executives are restricted to social measures, for example, social separating, travel boycott, and full lockdown in numerous urban communities. In this way, there is a critical requirement for the disclosure of counteraction and treatment methodologies for the COVID-19. It is recognized that the turn of events and assessment of an antibody may require somewhere around a year; besides, immunization improvement ought to be perceptive of generally wellbeing and administrative issues [11]. While the disclosure and assessment of another medication should take significantly longer, the utilization of a current medication (or compound going through clinical preliminaries) to treat COVID-19 (drug reusing) appears to be the quickest methodology, since these mixtures have either administrative endorsement as medications or have cleared security concentrates on that demonstrate a remedial potential [1], [12].

By and large, time is an indispensable element in the pandemic condition, so that, fast discovery, inoculation, and treatment strategies can fundamentally decrease mortality. All over again drug disclosure and improvement for lesser-referred to illnesses, for example, COVID-19 is exorbitant and monotonous. Subsequently, elective strategies, for example, the computational medication reusing approach can speed up the revelation of new medications. In such manner, a few pipelines have been presented for in silico drug repositioning against COVID-19 [13]. Recently, molecular docking as a famous bioinformatics strategy has been exceptionally viewed as the center of the most medication repositioning cycle to accomplish compelling medication possibility to battle COVID-19 [7].

As different antibodies and medications are going through clinical basics from one side of the planet to the next, drug reusing has been one of the successful frameworks taken by the researchers across the globe to draw out an appropriate solution for the destruction of the smart Covid. Against viral remedies, for example, chloroquine and hydroxychloroquine, used to treat stomach related disease and joint anguish autonomously, were upheld in the USA to treat SARS-CoV-2 patients. A piece of different solutions like remdesivir, actemra, and galidesivir are now going through clinical preliminaries, yet there has been no vaccination or steady medication at this point embraced by the FDA for the repugnance or treatment of SARS-CoV-2 [14], [15].

In this study, we have chosen nucleoprotein as a target in molecular docking to check which drug is best suited. Nucleoprotein packages the positive strand viral genome RNA into a helical ribonucleocapsid (RNP). It plays a fundamental role during virion assembly through its interactions with the viral genome and membrane protein M [16], [17]. This enhances the efficiency of sub genomic viral RNA transcription as well as viral replication. As nucleoprotein expects a huge part in viral replication, it would, in general, be an exceptional goal for the disclosure of novel drug against this protein, engaging the obstacle of viral development [18], [19].

## II. Proposed Methodology

### A. Protein and ligand selection

This study related to the drug for covid-19, we choose the protein that is present in severe acute respiratory syndrome coronavirus 2, that is, Nucleoprotein (PDB ID: 6wkp) is used as a target protein for molecular docking. For docking, five ligand/drug are selected to check which ligand is the best suited with nucleoprotein.

Firstly, nucleoprotein is retrieved from *RCSB PDB* website. To get pure protein, all ligands attached are removed using *Pymol* software and saved the processed file in .pdb format. Fig. 2 shows the structural view of nucleoprotein. Different ligands such as cepharanthine [20], favipiravir [20], IDX184 [20], remdesivir [20] and ribavirin [20] were then retrieved from the *PubChem website* and saved 3D conformation file in SDF format. Fig. 3 A shows the structural view of ligand cepharanthine in 2D, whereas Fig. 3 B represents 3D structural view. Similarly, Figures 4 to 7 represents the 2D and 3D structural views of favipiravir, IDX184, remdesivir and ribavirin, respectively. Table 1 demonstrates the detailed description of ligands.

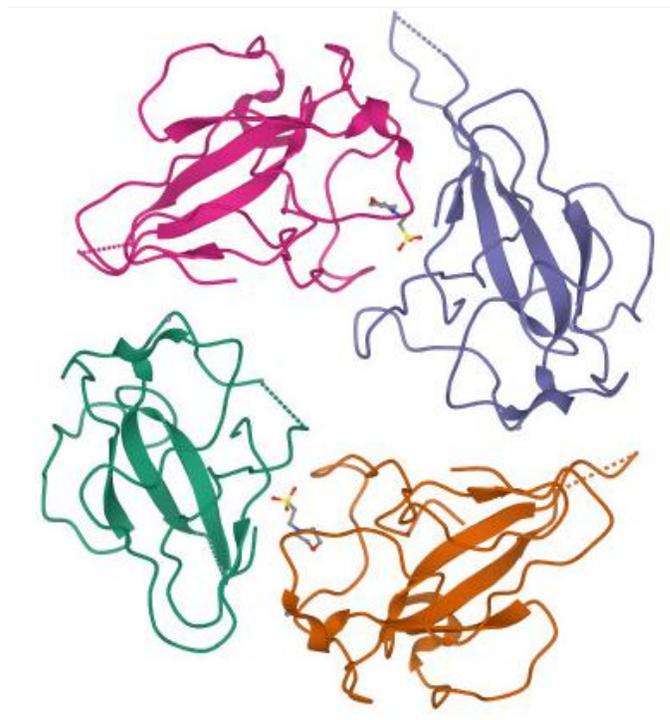

Fig. 2. Structural view of Nuclepprotein.

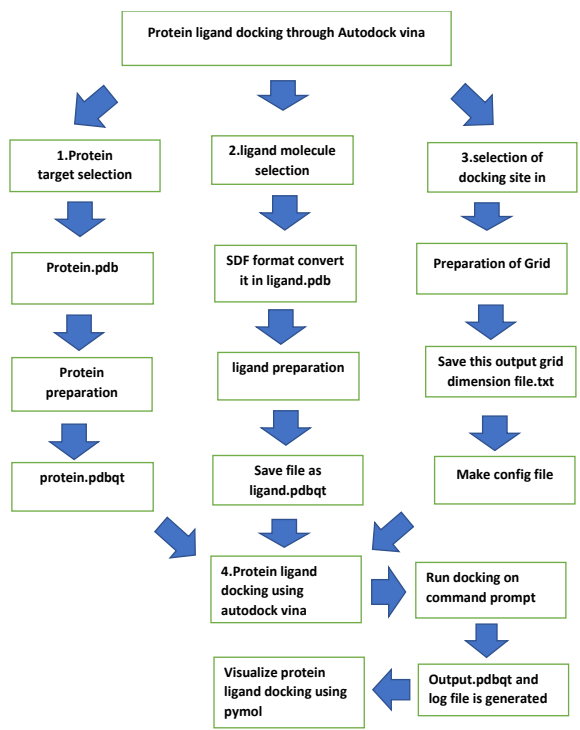

Fig. 1. Demonstrate the flow chart of docking procedure of different ligands/drugs using autodock vina software

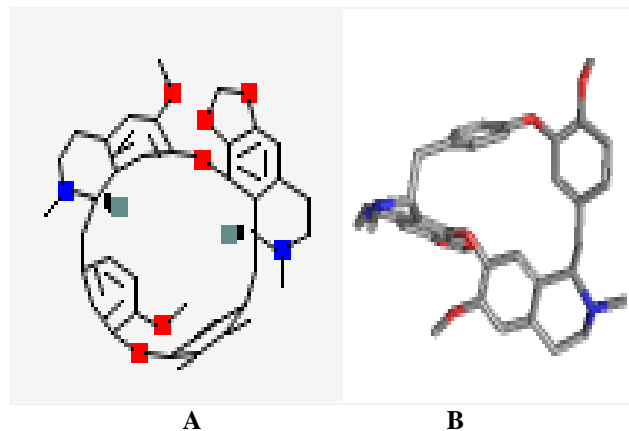

Fig. 3. (A) Ligand cepharanthine 2D and (B) Ligand cepharanthine 3D

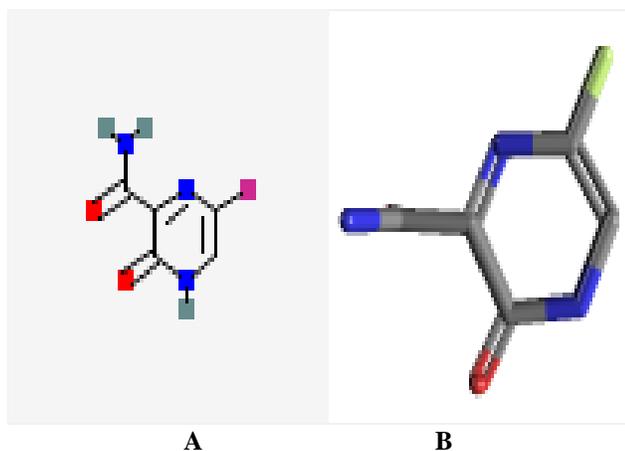

A B
Fig. 4. (A) Ligand favipiravir 2D and (B)Ligand favipiravir 3D

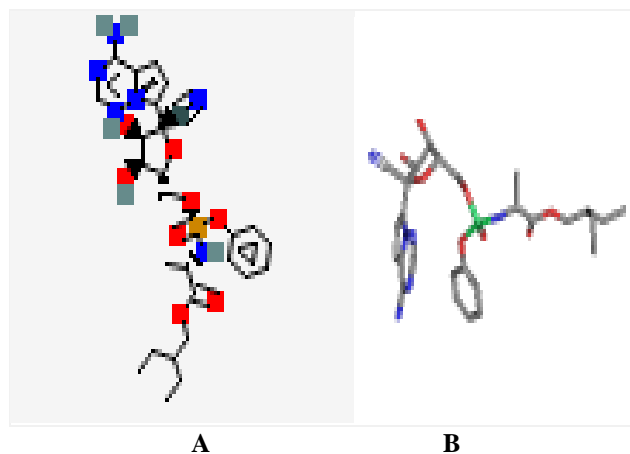

A B
Fig. 6. (A) Ligand Remdesivir in 2D and (B) Ligand Remdesivir in 3D

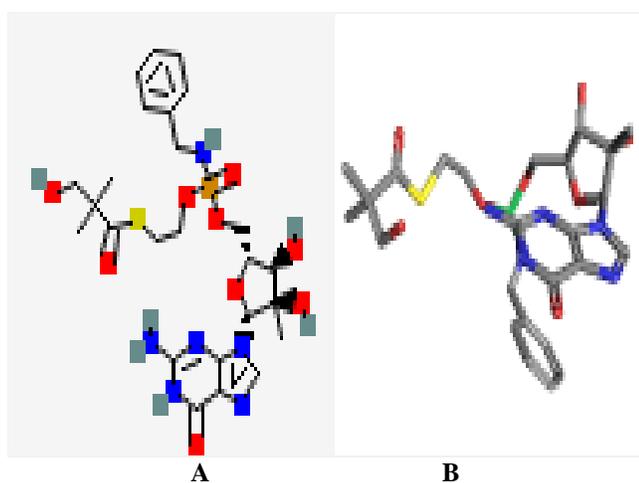

A B
Fig. 5. (A) Ligand IDX184 2D and (B)Ligand IDX184 3D

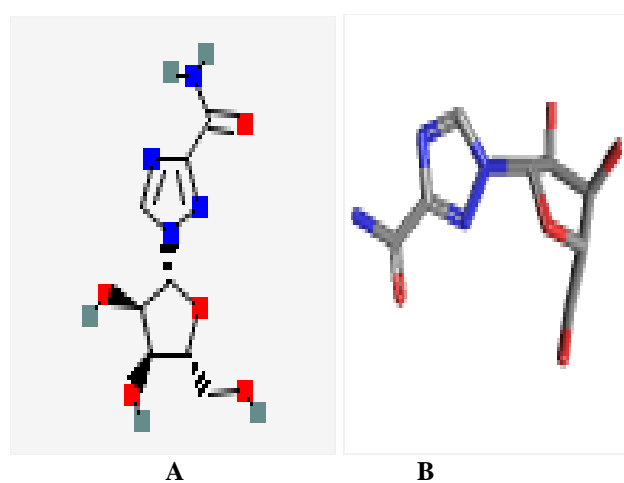

A B
Fig. 7. (A) Ligand Ribavirin 2D and (B)Ligand Ribavirin 3D

TABLE I. DESCRIPTION OF LIGANDS

| Sr. No | FDA UNII Code | Name | PubChem ID | Molecular Formula | CAS ID | Molecular weight |
|---|---|---|---|---|---|---|
| 1 | 7592YJ0J6T | Cepharanthine | 10206 | $C_{37}H_{38}N_2O_6$ | 481-49-2 | 606.7g/mol |
| 2 | EW5GL2X7E0 | Favipiravir | 492405 | $C_5H_4FN_3O_2$ | 259793-96-9 | 157.1g/mol |
| 3 | 4W44B4S9OC | IDX184 | 135565589 | $C_{25}H_{35}N_6O_9PS$ | 1036915-08-8 | 626.6g/mol |
| 4 | 3QKI37EEHE | Remdesivir | 121304016 | $C_{27}H_{35}N_6O_8P$ | 1809249-37-3 | 602.6g/mol |
| 5 | 49717AWG6K | Ribavirin | 37542 | $C_8H_{12}N_4O_5$ | 36791-04-5 | 244.2g/mol |

## B. Protein and ligand preparation

### 1) Protein preparation

To avoid the interference of water molecules in the pocket region, we used autodock tools that deleted water molecules from 3D structure of nucleoprotein. Then polar hydrogen atoms are added in the protein. Further, using Autodock vina, we saved the proceeded file in. pdbqt format. Fig. 8 shows the prepared view of nucleoprotein 6wkp.

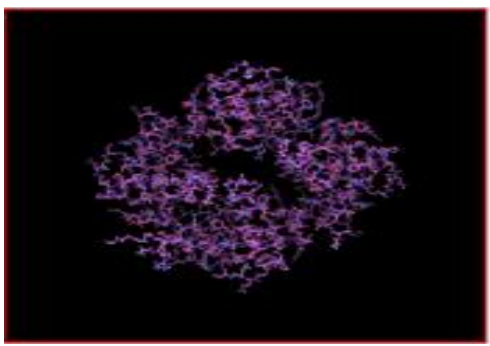

Fig. 8. Prepared view of nucleoprotein 6wkp

### 2) Ligand preparation procedure

Ligand file is converted in PDB format using Pymol software. First, SDF file of ligand is chosen, exported ligand molecule and save as PDB format. The ligand file is imported the auto dock tools. For ligand preparation click on ligand click on input then click on choose then dialog box appears. Choose ligand and then select molecule for autodock. Next step is to save as pdbqt file. Now we have the files cepharanthine.pdbqt and nucleoprotein.pdbqt. In this way, second step of docking is completed. Fig. 9 shows the prepared view of ligand cepharanthine. All other ligands favipiravir, IDX184, remdesivir and ribavirin are also prepared using the same procedure.

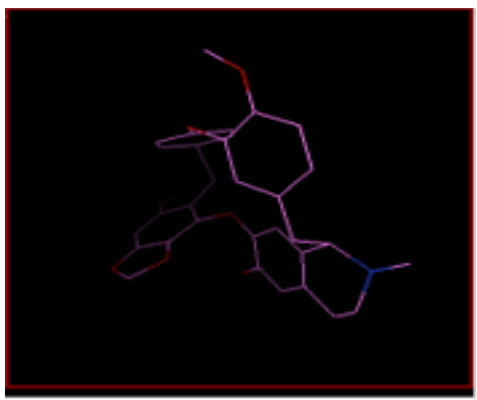

Fig. 9. Prepared view of Cepharanthine

## C. Selection of active site in protein

### 1) Preparation of Grid

Ligand in pdbqt format is drop in autodock tools. Corresponding protein and ligand view is shown in Fig. 10. The pocket site of protein is chosen for docking. The grid formation is chosen. The protein and ligand view produced using auto dock tools after choosing the protein as macromolecule is shown in Fig. 11. The active site of protein is found using the grid box settings. That is the region where docking takes place, and the dialog box shows the dimensions in grid.txt file. The grid view of protein is shown in Fig. 12.

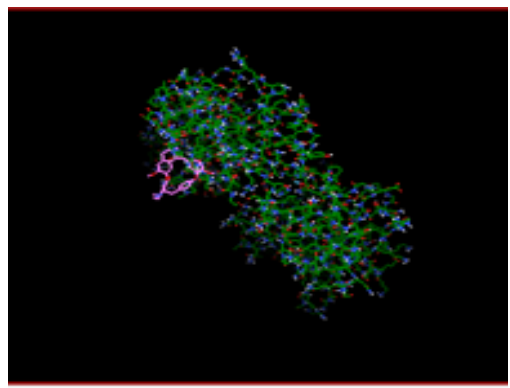

Fig. 10. Protein and ligand view in auto dock tools before docking

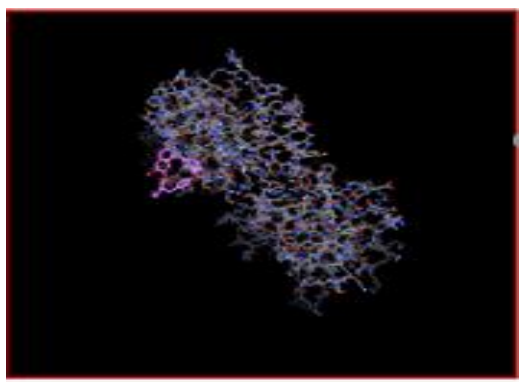

Fig. 11. Visualization of Protein as macromolecule and Ligand

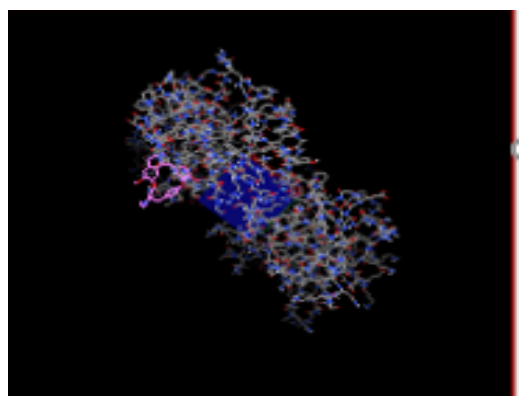

Fig. 12. Grid view of protein

## D. Protein ligand docking using autodock vina

For protein ligand docking, first, configuration file is created. It contains all necessary information that are required for docking. The docking process using autodock vina is executed in command mode. Once the docking is completed log and output.pdbqt file is generated. The Output.pdbqt generated file has ligand poses and corresponding log file shows the values of binding affinity in kcal/mol. Same docking procedure was performed for the docking of nucleoprotein with favipiravir, IDX184, remdesivir, ribavirin to check which one is the best suited drug for SARS-CoV-2 and their affinities is calculated in kcal/mol.

## III. RESULTS AND DISCUSSION

Docking of nucleoprotein with ligand cepharanthine from autodock vina gives output.pdbqt file and log file with 9 poses that have different values of binding affinity -9.4, -9.4, -9.3, -9.2, -8.6, -8.6, -8.4,- 8.3, - 8.1 in terms of kcal/mol. For good ligand docking performance, the lower value of binding affinity is better. Therefore, for cepharanthine compound, the lower binding affinity values for pose 1, 2 are found as a best docking pose. These results are highlighted in Fig. 13 A and B that show the best two poses of nucleoprotein 6wkp with ligand cepharanthine having the best binding affinity of -9.4 kcal/mol.

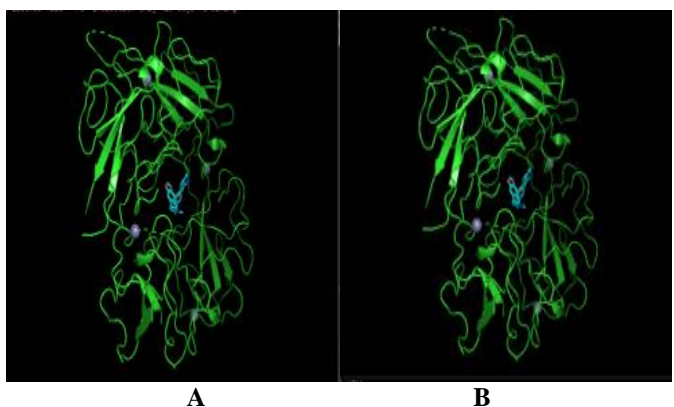

Fig. 13. Best docking poses in (A) and (B) of nucleoprotein with cepharanthine..

Docking of nucleoprotein with favipiravir from autodock vina gives output.pdbqt and log file that contain 9 poses with different values of binding affinity -6.1, -5.9, -5.7, -5.7, -5.6, -5.6, -5.5, -5.5, and -5.5 in terms of kcal/mol. The lower binding affinity of pose 1 is found as the best docking pose for favipiravir. Fig. 14 shows the best pose of nucleoprotein 6wkp with ligand favipiravir that has the minimum value of binding affinity -6.1 kcal/mol.

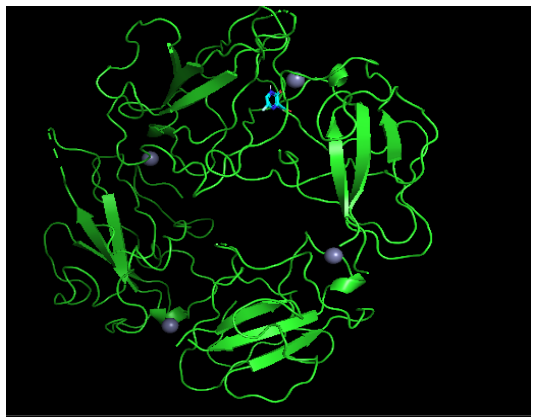

Fig. 14. Best docking pose of nucleoprotein with Favipiravir.

Docking of nucleoprotein with IDX184 from autodock vina gives output.pdbqt and log file that contain 9 poses having binding affinity values of -7.6, -7.4, -7.3, -7.3, -7.3, -7.2, -7.2, -7.1, and -7.1 kcal/mol. From this experiment, the lower value of binding affinity for pose 1 is found the best docking pose of IDX184. Corresponding Fig. 15 shows the best pose of nucleoprotein with ligand IDX184 with the best binding affinity of -7.6 kcal/mol.

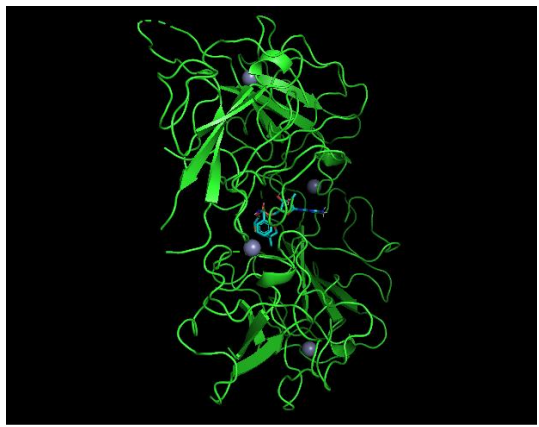

Fig. 15. Best docking pose of nucleoprotein with IDX184.

Further, docking of nucleoprotein with remdesivir from autodock vina gives 9 poses giving different values of binding affinity -8.5, -8.4, -8.3, -8.2, -8.2, -8.2, -8.2, -8.1, and -7.1 in terms of kcal/mol. The lower the binding affinity value for pose 1 is the best docking with remdesivir. Fig. 16 shows the best pose 1 of nucleoprotein 6wkp with ligand remdesivir that has the best binding affinity value of -8.5 kcal/mol.

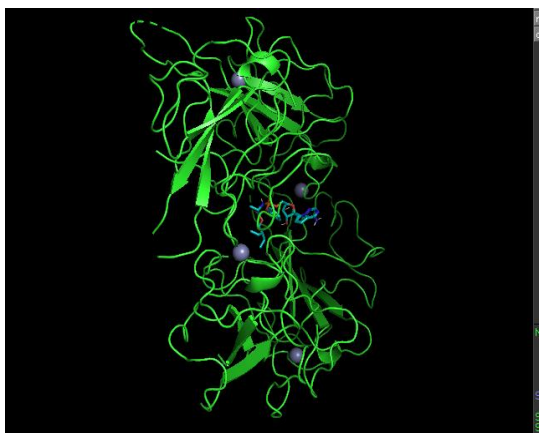

Fig. 16. Best docking pose of nucleoprotein with Remdesivir.

Finally, the docking of nucleoprotein with ribavirin including 9 poses that has nine values of binding affinity -7.1, -7.0, -7.0, -6.9, -6.6, -6.5, -6.4, -6.3, and -6.2. From this result, the lower value of binding affinity for pose 1 is found the best docking pose of ribavirin. Corresponding Fig. 17 shows the best pose of nucleoprotein with ligand ribavirin with the best binding affinity of -7.1 kcal/mol.

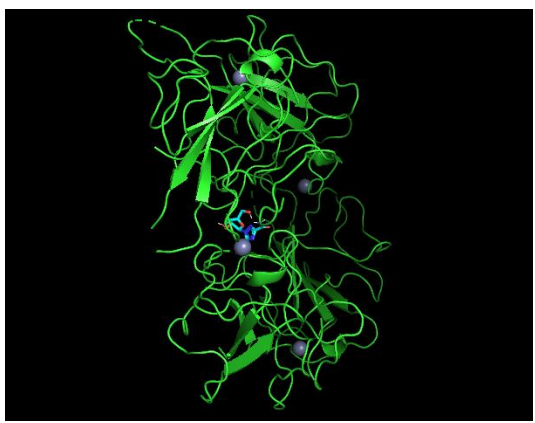

Fig. 17. Best docking pose of nucleoprotein with Ribavirin.

From this experiment, Table 2 described the summarized the best protein-ligand docking results in terms of binding affinity for their specific poses. This table demonstrates that ligand Cepharanthine has the best docking value with protein nucleoprotein (PDB ID: 6wkp) in terms of binding affinity-9.4 kcal/mol. Among other protein-ligands dockings. Other ligands such as favipiravir, IDX184, remdesivir, and ribavirin have shown lower docking results with nucleoprotein 6wkp.

TABLE II. COMPARISON OF BINDING AFFINITIES OF DIFFERENT LIGANDS WITH NUCLEOPROTEIN 6WKP

| Protein | ligand | Binding Affinity (kcal/mol) |
|---|---|---|
| Nucleoprotein 6wkp | Cepharanthine | **-9.4** |
| Nucleoprotein 6wkp | Favipiravir | -6.1 |
| Nucleoprotein 6wkp | IDX184 | -7.6 |
| Nucleoprotein 6wkp | Remdesivir | -8.5 |
| Nucleoprotein 6wkp | Ribavirin | -7.1 |

## IV. Conclusion

In this study, we explored the FDA approved drugs for recent pandemic of SARS-COV-2. We found that drug Cepharanthine is the best suited lead drug candidate against the nucleoprotein (PDB ID: 6wkp) of Covid-19. This protein is viral protein. Further, through the comparative studied, we found this drug is the most suitable candidate with other existing lead candidates such as favipiravir, IDX184, remdesivir and ribavirin. We found the best molecular docking parameter of binding affinity value -9.4, then other leading candidates. The therapeutic potential of Cepharanthine as an antiviral agent has been more important in combating COVID-19 caused by severe acute respiratory syndrome coronavirus-2 virus. Cepharanthine suppresses nuclear factor-kappa B activation, lipid peroxidation that are crucial to viral replication and inflammatory response. Against SARS-CoV-2 CEP predominantly inhibits viral entry and replication at low doses. We concluded that Cepharanthine drug can better combat against SARS-COV-2 nucleoprotein.

## Conflicts of Interest

The authors declare that they have no conflicts of interests.

## References

[1] C. N. Cavasotto and J. I. di Filippo, "In silico Drug Repurposing for COVID-19: Targeting SARS-CoV-2 Proteins through Docking and Consensus Ranking," *Molecular Informatics*, vol. 40, no. 1, Jan. 2021, doi: 10.1002/minf.202000115.

[2] Y. Yang *et al.*, "The deadly coronaviruses: The 2003 SARS pandemic and the 2020 novel coronavirus epidemic in China," *Journal of Autoimmunity*, vol. 109. Academic Press, May 01, 2020. doi: 10.1016/j.jaut.2020.102434.

[3] J. C. Rotondo *et al.*, "Sars-cov-2 infection: New molecular, phylogenetic, and pathogenetic insights. efficacy of current vaccines and the potential risk of variants," *Viruses*, vol. 13, no. 9. MDPI, Sep. 01, 2021. doi: 10.3390/v13091687.

[4] U. Anand *et al.*, "Potential Therapeutic Targets and Vaccine Development for SARS-CoV-2/COVID-19 Pandemic Management: A Review on the Recent Update," *Frontiers in Immunology*, vol. 12. Frontiers Media S.A., Jun. 30, 2021. doi: 10.3389/fimmu.2021.658519.

[5] E. Szliszka, Z. P. Czuba, M. Domino, B. Mazur, G. Zydowicz, and W. Krol, "Ethanolic Extract of Propolis (EEP) Enhances the Apoptosis- Inducing Potential of TRAIL in Cancer Cells," *Molecules*, vol. 14, no. 2, pp. 738–754, Feb. 2009, doi: 10.3390/molecules.

[6] Q. Wang *et al.*, "Longitudinal waning of mRNA vaccine-induced neutralizing antibodies against SARS-CoV-2 detected by an LFIA rapid test," *Antibody Therapeutics*, vol. 5, no. 1, pp. 55–62, Jan. 2022, doi: 10.1093/abt/tbac004.

[7] G. Arora, J. Joshi, R. S. Mandal, N. Shrivastava, R. Virmani, and T. Sethi, "Artificial intelligence in surveillance, diagnosis, drug discovery and vaccine development against covid-19," *Pathogens*, vol. 10, no. 8. MDPI, Aug. 01, 2021. doi: 10.3390/pathogens10081048.

[8] K. J. Downes *et al.*, "Return to School and COVID-19 Vaccination for Pediatric Solid Organ Transplant Recipients in the United States: Expert Opinion for 2021-2022," *J Pediatric Infect Dis Soc*, vol. 11, no. 2, pp. 43–54, Feb. 2022, doi: 10.1093/jpids/piab098.

[9] A. S. Bryant, "Major U.S. Pandemics a Century Apart: Preparedness and the Socioeconomic Impact of 1918 Influenza and 2020 COVID-19." [Online]. Available: https://digitalcommons.murraystate.edu/bis437/316

[10] V. Principal, L. Sow Kamaltai Jamkar Mahila Mahavidyalaya, I. Gollapalli Tejeswara Rao, U. Wamanrao Fule, and N. Dattatraya Totewad, *Editors Dr. Sangeeta Govindrao Avachar*. [Online]. Available: https://www.bhumipublishing.com/books/

[11] S. S. Jean, P. I. Lee, and P. R. Hsueh, "Treatment options for COVID-19: The reality and challenges," *Journal of Microbiology, Immunology and Infection*, vol. 53, no. 3. Elsevier Ltd, pp. 436–443, Jun. 01, 2020. doi: 10.1016/j.jmii.2020.03.034.

[12] S. Mahmood, T. Mahmood, N. Iqbal, S. Sabir, S. Javed, and M. Zia-Ul-Haq, "Traditional Chinese Medicines as Possible Remedy Against SARS-CoV-2," in *Alternative Medicine Interventions for COVID-19*, Springer International Publishing, 2021, pp. 63–109. doi: 10.1007/978-3-030-67989-7_3.

[13] Z. Madjd, S. Vafaei, M. Razmi, M. Mansoori, and M. Asadi-Lari, "The Lancet Infectious Diseases." [Online]. Available: https://ssrn.com/abstract=3569866

[14] M. S. Green *et al.*, "ASPHER statement on the need for a coordinated, professional approach to policy and planning for COVID-19 vaccination roll-out."

[15] C. K. Chang, M. H. Hou, C. F. Chang, C. D. Hsiao, and T. H. Huang, "The SARS coronavirus nucleocapsid protein - Forms and

functions," *Antiviral Research*, vol. 103, no. 1. pp. 39–50, Mar. 2014. doi: 10.1016/j.antiviral.2013.12.009.

[16] R. McBride, M. van Zyl, and B. C. Fielding, "The coronavirus nucleocapsid is a multifunctional protein," *Viruses*, vol. 6, no. 8. MDPI AG, pp. 2991–3018, Aug. 07, 2014. doi: 10.3390/v6082991.

[17] A. Jack *et al.*, "SARS-CoV-2 nucleocapsid protein forms condensates with viral genomic RNA," *PLoS Biology*, vol. 19, no. 10, Oct. 2021, doi: 10.1371/journal.pbio.3001425.

[18] S. K. M. Haque, O. Ashwaq, A. Sarief, and A. K. Azad John Mohamed, "A comprehensive review about SARS-CoV-2," *Future Virology*, vol. 15, no. 9. Future Medicine Ltd., pp. 625–648, Sep. 01, 2020. doi: 10.2217/fvl-2020-0124.

[19] S. Ahmad, "A Review of COVID-19 (Coronavirus Disease-2019) Diagnosis, Treatments and Prevention," *Eurasian Journal of Medicine and Oncology*, 2020, doi: 10.14744/ejmo.2020.90853.

[20] Singh, T. U., Parida, S., Lingaraju, M. C., Kesavan, M., Kumar, D., & Singh, R. K. (2020). Drug repurposing approach to fight COVID-19. *Pharmacological Reports*, 72(6), 1479-1508.